\documentclass{acm_proc_article-sp}
\usepackage{hyperref}

\begin{document}

\title{Recommending More Efficient \\ Workflows to Software Developers} 
%

\numberofauthors{1}
\author{
%
%
\alignauthor
Dylan Bates\\
       \affaddr{Coker University}\\
       \email{dylan.bates@coker.edu}
}
\date{\today}

\maketitle
\begin{abstract}
Existing recommendation systems can help developers improve their software development abilities by recommending new programming tools, such as a refactoring tool or a program navigation tool. However, simply recommending tools in isolation may not, in and of itself, allow developers to successfully complete their tasks. In this paper, I introduce a new recommendation system that recommends workflows, or sequences of tools, to developers. By learning more efficient workflows, the system could make software developers more efficient.
\end{abstract}



\section{Research Problem \\ and Motivation}
Software developers use tools in order to make programming easier, by reducing development time and improving software quality. However, software developers only use a small fraction of the total number of tools available to them~\cite{Murphy-Hill:1}. Several programs combat this, by recommending tools to the user. Some of these programs work by simply recommending the most popular tools in an integrated development environment that the user does not make use of~\cite{Linton:1}. Others work by comparing a user's tool frequencies to the entire user population~\cite{Matejka:1}. Exposure to more tools encourages the developer to use these tools to reduce development time and ultimately increase software quality.

The sequential use of several tools characterize a software developer's workflow, which Linton and colleagues define as ``one or more plans for attaining a goal,'' which ``ultimately decompose to a sequence of actions''~\cite{Linton:1}. In this paper, I interpret this as a series of tools used one after another, in order to accomplish a specific, unique, and efficient task. I propose building on prior work that recommends tools, by instead recommending more efficient workflows.

Consider the following example of an inefficient workflow.
Suppose Evelyn is working in the Eclipse development environment\footnote{\url{https://www.eclipse.org}}, repeatedly using the \texttt{Find References} command on several methods in a call chain. In this example, the task she is trying to accomplish is walking up a call hierarchy.  This is an example of an inefficient workflow, because she could accomplish the same task by using the \texttt{Call Hierarchy} tool, which takes fewer steps than using \texttt{Find References} repeatedly. Most workflows use several tools, such as \texttt{Copy/Paste} to duplicate and move text and \texttt{Organize Imports/Format/Save All} to clean up code.


The insight of this paper is that I can use some of the same ideas to recommend workflows as were previously used to recommend tools.

My paper makes three contributions: it introduces a system that recognizes the most common workflows comprised of $n$ tools; it demonstrates how $n$-tool workflows (called \textit{$n$-flows}) can then be used to recommend a workflow to a software developer; and it explains how a recommendation system could implement this technique of recommending workflows, potentially increasing developers' efficiency and productivity.

\section{Background and Related Work}
\label{alg}
The previous work in this field recommends single tools (which I define as $1$-flows) to a user. Existing systems do this via one of several algorithms, such as content-based filtering~\cite{Matejka:1, Murphy-Hill:1}, collaborative filtering~\cite{Matejka:1, Murphy-Hill:1}, and most popular~\cite{Linton:1}.
 
The main difference between each of these methods is the way they generate the items to recommend. These methods have already been used by Matejka and colleagues~\cite{Matejka:1}, Murphy-Hill and colleagues~\cite{Murphy-Hill:1}, and Linton and colleagues~\cite{Linton:1} to recommend tools. One could feasibly extend their methods to recommend workflows to developers.

These methods are focused on recommending single tools that will potentially increase the productivity of a developer. The problem with existing implementations is that given out of context, some tools do not constitute a complete task, and are therefore difficult for a user to implement effectively. This problem was addressed by Viriyakattiyaporn and Murphy~\cite{Viriyakattiyaporn:1} when they implemented a system called \textit{Spyglass}, which attempts to recognize inefficient navigational workflows and suggests tools to aid program navigation as a developer works. Unfortunately, Spyglass is limited in that can only recommend single navigational tools, instead of potential workflows that users could implement.

\section{Approach and Uniqueness}
I went about the task of identifying the most common workflows by analyzing about 23 million time-stamped tool uses from 4308 Eclipse users, collected from the Eclipse Usage Data Collector\footnote{\url{http://www.eclipse.org/org/usagedata/index.php}}. A total of 700 unique tools were used. I attempted three ways to discover the most common $n$-flows, each of which has its own advantages and disadvantages.

These methods consisted of a Top-$K$ Sequential pattern mining algorithm called TKS~\cite{Fournier:1}, sorting an $n$-dimensional matrix, and using a \texttt{Map} to map $n$-flows to the number of uses. Each implementation worked well for different sized datasets and numbers of dimensions.

Through the use of all three algorithms, I was able to determine the top $n$-flows for $n<5$. I was then able to use Linton and colleagues' \textit{most popular} algorithm~\cite{Linton:1} to recommend workflows to any given user in the set.




By looking at $n$-flows instead of single tools, this system is able to recommend entire workflows to a user that they may not otherwise discover on their own. For example, if Evelyn is new to Eclipse, and using an existing recommendation system to learn new tools, it will recommend a single tool to her (for example, \texttt{Copy}). Without the proper documentation for the tool, Evelyn may be lost. Certain tools, when used make no observable changes to the screen; she will find that nothing apparent happens. Had Evelyn been recommended a workflow containing the tool (in this case \texttt{Copy/Paste}), it may be more clear what the workflow accomplishes. In this case, the context for use is then clear. This method can eliminate the \textit{discoverability barrier}~\cite{Murphy-Hill:1} presented by Murphy-Hill and colleagues, as developers will learn new tools as well as more efficient workflows. 

When cleaning the data, I had to prune the dataset to remove all repeated uses of tools, such as deleting an entire line one character at a time, or saving seven times before doing anything else. Before doing this, the top five workflows for $n<5$ all consisted of repeated uses of a single tool; for example \texttt{Delete/Delete/Delete} or \texttt{Save/Save/Save/Save}. These were removed from the results.

\section{Results and Contribution}
My research suggests that while there are several common workflows that Eclipse users implement in order to accomplish a task, there are many others that do not appear to serve a specific purpose. I collected the top 100 $n$-flows for $n<5$,\footnote{Available at \url{http://goo.gl/nRKMv3}} and was able to generate 209 workflows and individual tools to recommend to a user in the dataset. I found that many of the $n$-flows I discovered did not meet my definition of a workflow, as they did not appear to accomplish a specific, unique, or efficient task. Due to this, most should not be recommended in a system that recommends more efficient workflows.

While $2$-flows occurred most frequently, they are also the most trivial. For example, the most common $2$-flow was \texttt{Copy/Paste}, which $>99\%$ of all users in the dataset used. The most common $3$-flow was \texttt{Paste/Copy/Paste}. The most common $4$-flow was \texttt{Copy/Paste/Copy/Paste}. Each of these $n$-flows is a subset of an $(n+1)$-flow, which was found to occur throughout the $n$-flows discovered.  Interestingly, the top eight $n$-flows for $n<5$ consisted of various permutations of the following tools: \texttt{Copy}, \texttt{Cut}, \texttt{Delete}, \texttt{Paste}, and \texttt{Save}. In fact, most of the $n$-flows I discovered consisted of these five tools, as well as simple text editing tools, such as \texttt{Go To Line Start}, \texttt{Select To Line End}, \texttt{Go To Previous Word}, and \texttt{Select Next Word}.

\subsection{Limitations}
The major limitation that I faced was the noise found in mining common workflows, which makes recommending more efficient workflows difficult. That is to say, the vast majority of $n$-flows found do not appear to complete a task. Also, the most common $n$-flows found are redundant, in that they are often a subset of another common workflow. Additionally, the most frequently occurring $n$-flows consist of tools that nearly everybody uses, such as saving, moving the cursor, and selecting text, which means they would not be as effective for recommendations. In a recommendation system, workflows would need to be recommended that are necessary, efficient, and accomplish a task.

\subsection{Future Work}
Building on the results presented here, I can use the idea of recommending workflows to create a functioning recommendation system based on the algorithms mentioned in Section~\ref{alg}. 
Theoretically, it could be possible to go through these $n$-flows and recommend only those that meet the definition of a workflow, as defined earlier. Then, a recommendation system could be made that only recommends relevant workflows to users. Finally, a study could be carried out to determine the effectiveness of recommending workflows. Only then could the insight of this paper be confirmed: that learning more efficient workflows makes software developers more efficient.

%
\bibliographystyle{abbrv}


\balancecolumns
\end{document}